# Terahertz orbital angular momentum modes with flexible twisted hollow core antiresonant fiber

Alessio Stefani,[1,2,a)] Simon C. Fleming,[1] and Boris T. Kuhlmey[1]

[1]*Institute of Photonics and Optical Science (IPOS), The School of Physics, The University of Sydney, NSW 2006, Australia*

[2]*DTU Fotonik, Department of Photonics Engineering, Technical University of Denmark, 2800, Kgs. Lyngby, Denmark*

THz radiation is more and more commonplace in research laboratories as well as in everyday life, with applications ranging from body scanners at airport security to short range wireless communications. In the optical domain, waveguides and other devices to manipulate radiation are well established. This is not yet the case in the THz regime because of the strong interaction of THz radiation with matter, leading to absorption, and the millimeter size of the wavelength and therefore of the required waveguides. We propose the use of a new material, polyurethane, for waveguides that allows high flexibility, overcoming the problem that large sizes otherwise result in rigid structures. With this material we realize antiresonant hollow-core waveguides and we use the flexibility of the material to mechanically twist the waveguide in a tunable and reversible manner, with twist periods as short as tens of wavelengths. Twisting the waveguide, we demonstrate the generation of modes carrying orbital angular momentum. We use THz time domain spectroscopy to measure and clearly visualize the vortex nature of the mode, which is difficult in the optical domain. The proposed waveguide is a new platform offering new perspectives for THz guidance and particularly mode manipulation. The demonstrated ability to generate modes with orbital angular momentum within a waveguide, in a controllable manner, will be beneficial to both fundamental, e.g. matter-radiation interaction, and applied, e.g. THz telecommunications, advances of THz research and technology. Moreover, this platform is not limited to the THz domain and could be scaled for other electromagnetic wavelengths.

[a)] Author to whom correspondence should be addressed. Electronic mail: alessio.stefani@sydney.edu.au.

**INTRODUCTION**

The rapidly increasing availability of sources and detectors has brought the field of THz radiation, also known as millimeter waves, from an interesting physics regime to the real world, with applications in security, telecommunications, characterization of biological and solid-state matter, and many others.[1,2] These various applications require, on top of generating and detecting the desired radiation, means of transporting and manipulating the fields.

In the context of guiding THz radiation, there has been a large effort in realizing waveguides which are at the same time small, flexible and low loss. The size of the waveguides is constrained by the wavelength to be guided and therefore THz waveguides must have millimeter to centimeter sized cross sections, which makes them very rigid. Furthermore, the material providing the confinement strongly affects the loss of the waveguides; unlike at optical wavelengths, there is no material with very low loss in the THz region, so that the use of hollow waveguides is necessary. A variety of types of hollow waveguides have been studied for THz, which may be separated into metallic, dielectric and hybrid waveguides (including metamaterial waveguides).[3,4,5,6,7,8,9] Metallic hollow waveguides can have subwavelength cross-section and have relatively low loss. Those characteristics arguably make them the current best option for THz waveguides. Metamaterial hollow waveguides offer the potential for control of the radiation,[5] can be subwavelength[6] and potentially low loss.[7] However, both complexity of fabrication and the relative lack of research in this field leaves doubts about their real potential. Dielectric THz hollow waveguides are the scaled version of their more successful optical counterpart and have been investigated in the form of Kagome structures,[10,11] simple capillaries,[12] antiresonant structures,[13] and several other variations.[8,9] Although some of these have the possibility of achieving excellent guidance, the relationship between loss and size has meant that realizations of this approach are large rigid structures that were of little practical use. Leaving aside the rigidity issue temporarily, there is one class of hollow dielectric waveguides that deserves more detailed consideration, mostly because of the greatly expanding recent interest and results in the optical domain: antiresonant fibers.

Antiresonant fibers[14] have been, in the last five years, dominating the panorama of hollow core fibers. The reason is that a very simple structure, i.e. a circular array of capillaries, allows guidance of very large bandwidths with very low loss almost independently of the absorption of the material used for the structure. Moreover, by tuning the few parameters of this structure it is possible to obtain single mode operation in large cores[15] and to tune the bending loss of the structure.[16] In the context of THz, the realization of such fibers allows reducing the overall structure size compared to bandgap or Kagome structures, controlling the modal properties and further reducing the waveguide loss with any material. Although interesting, this is not sufficient unless the waveguide can be made usable in practical applications. The enabling factor behind this work is the use of a novel material for waveguides, namely polyurethane, the Young's modulus of which is 2 to 3 orders of magnitude lower than conventional waveguide dielectrics and allows even centimeter sized tubes, rods and structures to be bent with radius 10 times the diameter. This means a 1 cm diameter fiber can be bent in a 10 cm radius circle (an example is shown in Fig. S1 of the supplementary material). This unique property allows the fabrication of waveguides with large cross sections that are still flexible, and allows for further mechanical manipulations of the fibers. The specific manipulation we will exploit in this paper is twisting of the structure in a controlled and reversible way.

The idea of twisting fiber structures is very interesting because it enables creating modes with an orbital angular momentum (OAM) [17,18,19] and controlling the state of polarization and the fiber's optical activity.[20,21] It should be kept in mind that in order to obtain any effect, the fiber to be twisted cannot be cylindrically symmetric. There have been interesting recent advances in this research by twisting photonic crystal fibers (PCFs). So far, there have been two ways to give PCFs a twist: by post processing with a $CO_2$ laser[17,18] or by twisting the preform during fiber drawing.[18,22,23] Both methods result in a permanent twist of the fiber and the twist occurs at high temperatures where the material viscosity is low. When in a solid state, mechanical, and therefore reversible, twist of silica fibers cannot be used to produce twist rates high enough to achieve the desired properties,[17,18,19,20,21,22,23] because of the inherent material stiffness. Moreover, even when fabricated with a permanent deformation, these twisted structures have been realized only in the optical domain and only theoretically suggested in the THz, once more because of the combination of size and necessary deformation required to achieve structures showing interesting properties.

The interest behind being able to realize modes that have orbital angular momentum derives from the plethora of applications this class of modes has found in 25 years of investigation,[24,25,26,27] including but not limited to, imaging and microscopy,[28] quantum photonics and entangled states,[29] optical tweezers,[30] and optical communications.[31] Modes carry orbital angular momentum when their phase structure is rotating. Assuming an $e^{i\omega t}$ time dependence, they have an optical vortex along their axis and an azimuthal phase dependence of the form $e^{-i\ell\phi}$, with $\ell$ being a positive or negative integer and $\phi$ the angular coordinate, and they carry a momentum of $\ell\hbar$ per photon. The helical phase front and the azimuthal $2\pi\ell$ phase change require a phase singularity in the middle which leads to a central intensity null.[27] Because of the large interest involved, there have been many proposals on how to generate OAM modes, including phase plates,[32] diffractive elements using holograms[33] or spatial light modulators,[34] mode conversion with cylindrical lenses,[24] metasurfaces,[35,36] microbend gratings in fibers[37,38] and twisted PCFs.[18] The generation of OAM for THz radiation has been done similarly, by using some of the techniques also used for optical beams.[39,40,41,42,43] However, the ability to generate OAM within a waveguide in a tunable manner is very appealing both in the optical and especially in the THz domain where this combination was not possible before.

In this paper we report the realization of a simple, hollow core antiresonant waveguide, which is flexible and opens up a new class of THz waveguides. The flexibility of the waveguide is used to twist the fiber along its axis and to create THz vortex modes possessing orbital angular momentum, as represented in Fig. 1. THz time domain spectroscopy (TDS) is used to measure the vortex nature of OAM modes.

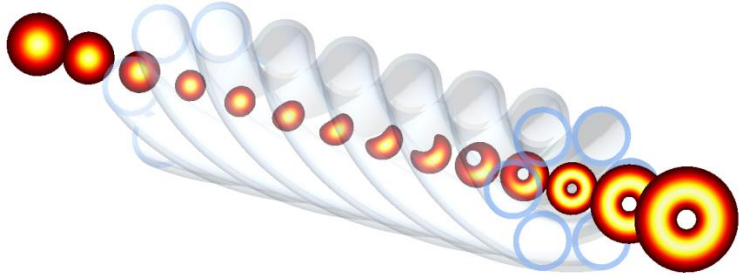

Fig. 1. Schematic of the proposed process. A fundamental-like mode is coupled into the fiber which is mechanically twisted and therefore allows for coupling from the fundamental mode to an OAM-possessing mode.

## FLEXIBLE HOLLOW CORE WAVEGUIDE

Polyurethane (PU) has a Young's modulus orders of magnitude lower compared to silica and can withstand elongations of 600%,[44] making it effectively a rubber-like elastic material. The ability to fiber-draw PU and the possibilities to apply its property to elastically deform have been recently demonstrated in the realization of a tunable metamaterial.[45] These mechanical properties make PU a perfect candidate for realizing a structure that can be twisted mechanically. The choice of a hollow core structure is due to both the poor optical properties of polyurethane and to the potential of this structure to be scaled at any wavelength independently from the material properties, in particular loss. As already mentioned, of the class of hollow core structures, the antiresonant tube lattice structure is the simplest and the most obvious choice. The proposed fiber is realized by six PU tubes arranged in a circle, as shown in the inset Fig. 2(a) and the schematics in Figs. 1 and 3(a). The tubes are commercially available with large outer sizes and are fiber-drawn to about 3 mm in outer diameter. The fiber structure is held together by gluing the tubes to a plastic disc on each end while the middle is free standing. The thickness of the disc is 5 mm and a hole was drilled to size to accommodate the tube lattice structure. While such a structure can be drawn to smaller dimensions if necessary, the structure was suitable without the necessity of further scaling for testing the principle in the THz regime. The resulting fiber core has a 3 mm diameter, and the capillaries have thickness of about $a$ =400 µm. Using phase resolved propagation measurements in 3 and 5 mm PU samples we calculated a refractive index of $n$ =1.6 at 0.3 THz. This yields resonances at frequency multiples of $f_R$ =300 GHz, calculated by using the analytical expression for antiresonant fibers with an air-core:[14]

$$f_R = \frac{c}{2a\sqrt{n^2 - 1}}, \tag{1}$$

where $c$ is the speed of light.

The fiber was characterized in a THz time domain spectroscopy system.[2] Figure 2(a) shows the transmitted spectrum through 10 cm of fiber for the frequency range 0.1 - 1.1 THz. As expected, bands of high transmission are visible. The minima of the transmission correspond quite well to the analytical calculation of the tube walls' resonances (red dashed lines in Fig. 2(a)). Similar transmission curves, with identical resonance limited bands, were obtained with fibers 12, 14.5, 17.5 and 20 cm long. However, because of the size of the waveguide and realization of the structure (how the tubes are kept in place) cutback measurements to obtain propagation losses were not possible: when comparing the transmission between the various sample lengths, the coupling efficiency played a significant role and a comparison between them leads to errors larger than the actual values. This is also due to the fiber being multimoded in a good part of the transmission window. As a reference, we measured ~10 dB loss including coupling and propagation around 0.8 THz for the 10 cm fiber. Simulations of the structure using a commercial finite element solver (COMSOL) show minimum propagation loss approaching 0.1 dB/m for the fundamental mode (Fig. 2(b)). For information about the loss of high order modes see the relevant section in the supplementary material. The simulated structure has core diameter of 3.2 mm and capillaries with 3 mm diameter, 400 µm wall thickness and refractive index of 1.6. The loss of PU is set to a constant value of 1 dB/cm for all frequencies which is the measured value of the bulk material at 0.5 THz. While such low simulated losses demonstrate the potential of this fiber, they are only found over a relatively narrow band and assume a perfect structure. It is instead reasonable to consider the lowest loss to be around 1-2 dB/m for the fundamental mode in the transmission bands of interest to this investigation. Note that loss could be further reduced with thinner and more widely spaced capillaries. Effective single mode behavior could be achieved by carefully choosing the ratio of the capillaries and core diameter[15] or by twisting the structure.[23]

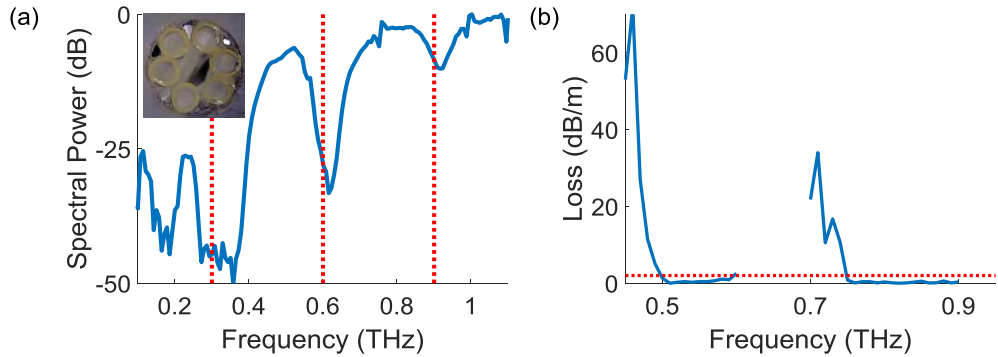

Fig. 2. (a) Normalized measured transmission through 10 cm of antiresonant PU fiber. A photograph of the cross section of the fiber is shown in the inset. The red dotted lines indicate the expected frequencies of the transmission minima calculated with Eq. (1). Data is normalized to the maximum measured power. (b) Simulated propagation loss of the fundamental mode. The red dotted line marks 2 dB/m loss.

## TWISTED FIBER FOR OAM GENERATION

The realized structure and its flexibility (also illustrated in Fig. S1 of the supplementary material) give us the possibility to investigate the effects of twist on the guided modes. The fiber was mechanically twisted, by keeping one end fixed and rotating the other. The amount of twist could be tuned and the twist is reversible. The twist we could apply was limited by the structural stability of the tube assembly: beyond a certain amount of twist either the adhesive would fail and the tube bundle dismantle or one of the tubes would slip into the core. For the lengths involved in this experiment, i.e. 10 to 20 cm, around 8 cm twist period was the shortest achievable reliably. Longer fibers allowed shorter twist periods, probably because of the force on the adhesive being lower. Figure 3(a) shows side images of the fiber during the twisting and un-twisting process and schematics of the twisted fiber.

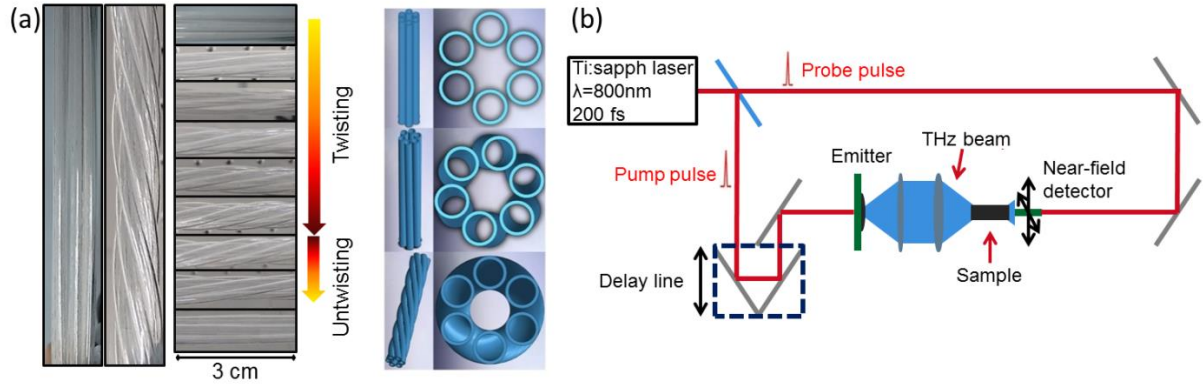

Fig. 3. (a) Various twisting and un-twisting steps and schematics of the twisted fiber offering a view of the waveguide when twisted. For information about the transmission of the fiber when twisted refer to the dedicated section in the supplementary material. (b) Schematic of the near-field raster scan THz TDS set-up.

In order to measure the modal content, near-field raster scan measurements were performed at the output end of the waveguide. The great advantage of THz TDS is that it allows us to acquire information on the intensity, electric field and phase of the modes and also it allows analyzing their time evolution. With this technique, the amplitude of the THz electric *field* (not its envelope or intensity) is directly measured as a function of the time delay between the pump and probe pulses generating and detecting the broadband THz pulses. Therefore, the time delay allows sampling of the THz electric field in time. The measured amplitude of the electric field can be mathematically processed with a Fourier transformation to obtain the complex electric field as a function of frequency. Accessing the complex field is necessary for obtaining full information about modes/fields where the phase of the electric field is a characteristic feature, as it is for modes carrying OAM. A schematic of the measurement system is shown in Fig. 3(b). The scan was performed on a 2.4x2.4 mm area with lateral steps of 75 μm, at distance of about 50 μm from the end-facet and scanning 64 ps time delay, resulting in a frequency resolution of 15 GHz. Photoconductive antennas where used for the THz generation (EKSPLA EMT-08) and detection (near-field probe TeraSpike TD-800-X-HR). The emitter generates a linear polarization (in the *x* direction), and the detection antenna was aligned to be sensitive to the field component aligned in the same direction as that of the excitation only. This limitation did not affect the intended ability to characterize orbital angular momentum modes. It did however hinder our ability to explore some of the expected effects of the twist on the polarization.

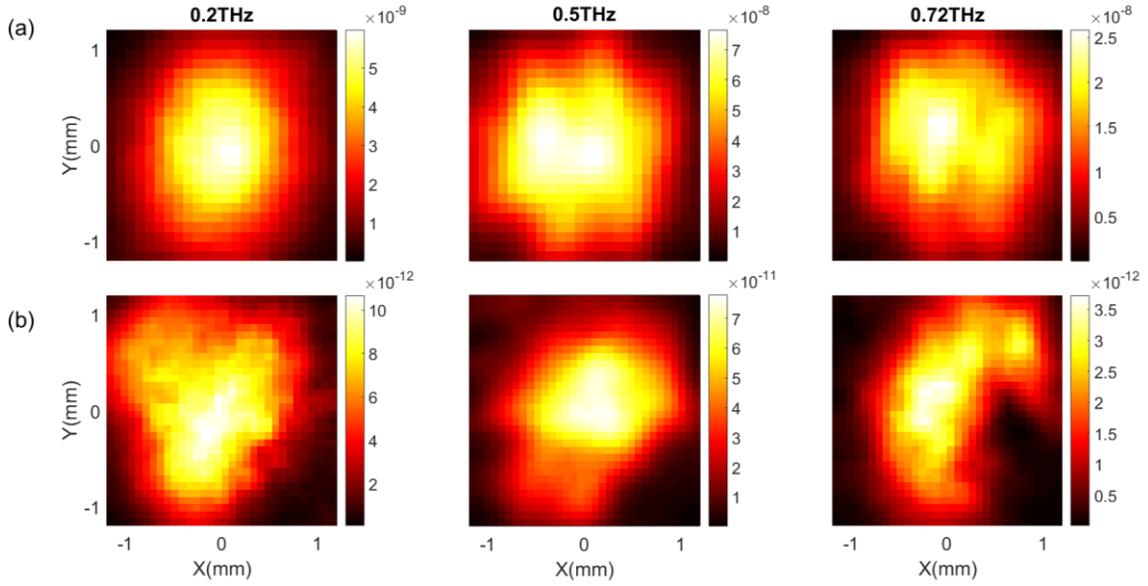

Fig. 4. Mode intensity profile $|E_x|^2$ at the end facet of (a) a 17.5 cm long untwisted fiber and (b) the 10 cm long fiber twisted with a 10 cm twist period in the 3 main transmission bands. The color scale is used as an indication to be able to compare mode intensities within the same measurement; but should not be used to compare intensities of different measurements as the values are strongly dependent on the measurement parameters.

The measured intensity of the *x*-component of the modal fields (one for each transmission window 0.2, 0.5 and 0.72 THz) of the 10 cm long fiber twisted with a 10 cm period are shown in Fig. 4(b). In the lowest two bands the intensity is consistent with fields mostly composed of the fundamental mode, while the fields in the 0.72 THz band shows a different profile, consistent with a superposition of modes, both in the twisted and untwisted case. When twisted a central singularity appears in the field distribution. Such an intensity profile is not surprising given the helicoidal fiber supports vortex modes with a central zero in the intensity. The analysis of the electric field distribution and its phase profile confirms the mode possesses orbital angular momentum having a 2π spiral phase front, as shown in Fig. 5. It should be noted that the intensity profiles and electric field distributions for the different frequencies within each transmission band are similar, implying that OAM is generated at every frequency between 0.6 and 0.9 THz. The electric field and phase distribution measured in lower frequency bands do not show significant variations from the fundamental mode, however as we did not measure the cross polarization it is possible that the polarization state has changed between input and output.

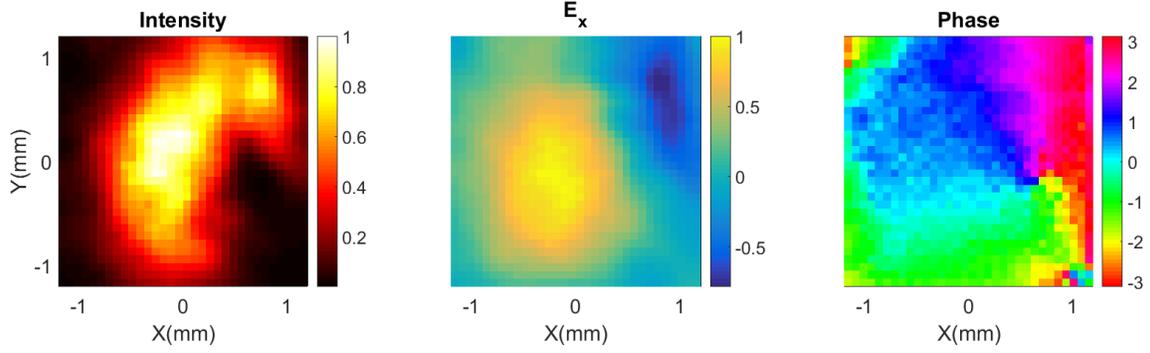

Fig. 5. Intensity profile $|E_x|^2$, x-component of the electric field, and phase measured at 0.72 THz.

Taking advantage of the time resolved nature of the measurement performed and using a windowed Fourier transform (time window of 26 ps), we obtain the time evolution of the various field components at selected frequency bands. A video with the evolution of the intensity, electric field and phase measured at 0.72 THz is included as a media file (Fig. 6). The electric field has two lobes opposite in sign spiraling around the fiber axis with time. Similar information is obtained by the evolution of phase, showing the rotating phase structure typical of modes with OAM. Figure 7 shows the isophase wavefronts evolution in time for the case with OAM (0.72 THz) and, for comparison, without OAM (0.5 THz). This measurement is a clear representation of the vortex nature of the field. The spacing between 2 events with the same phase corresponds to one propagation wavelength travelling at the phase velocity. The intensity profile (Fig. 6 and video) clearly shows the pulsed nature of the beam, with a much longer time duration compared to a single optical cycle.

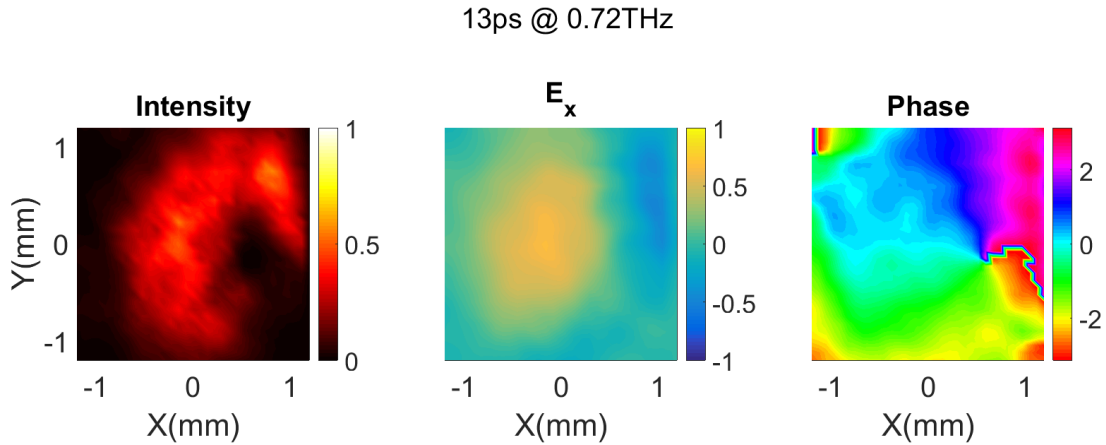

Fig. 6. Time evolution of the intensity profile $|E_x|^2$, x-component of the electric field, and phase at 0.72 THz (multimedia view).

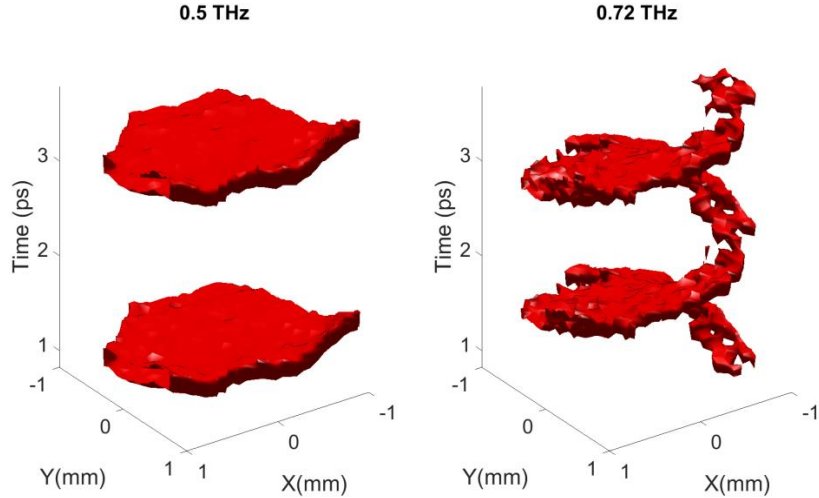

Fig. 7. Isophase evolution of the electric field as function of time at 0.5 THz and 0.72 THz

**DISCUSSION – NUMERICAL INVESTIGATION**

The field generated and measured clearly has OAM, but the appearance of it is not that of a clean integer OAM mode, for which the phase singularity would be centered, the intensity profile symmetric, and the phase profile radially uniform.

In order to better understand the process of OAM mode generation in this fiber, finite elements numerical simulations using a helical coordinate system transformation[17,18,21] were performed. In all the simulations reported in the paper a mode is considered to be guided in the core when more than 50% of the square of the electric field is contained in the circle inscribed within the capillaries. In the lower frequency transmission bands (0.2 and 0.5 THz) the fiber only supports one core mode, which does not have OAM in the twist rates range used in the experiments. Higher twist allows modes carrying OAM to be supported, but are quite lossy. This explains why we do not measure OAM modes in the lower bands. In the 0.6-0.9 THz transmission band, the straight fiber supports multiple guided modes, which allows guidance of OAM modes if properly excited, as also previously numerically reported for a similar structure at a similar frequency.[46] When twisted with 10 cm period, the fiber still supports the fundamental mode, although the degeneracy in polarization is lifted, as well as other modes all carrying OAM. It should be remembered that in circularly symmetric optical fibers there are 4 possible modes that can carry OAM for $\ell=1$, consisting of the $LP_{11}$ manifold: $TE_{01}$, $TM_{01}$, $HE_{21}^{odd}$, $HE_{21}^{even}$.[38] Modes with similar electric field distributions to all those modes are supported in the fiber. Figure 8 shows effective refractive indices of the modes in the fiber at 0.72

THz as a function of twist and some of the modes' profiles and phase. For information on the propagation loss of these modes see the relevant section in the supplementary material.

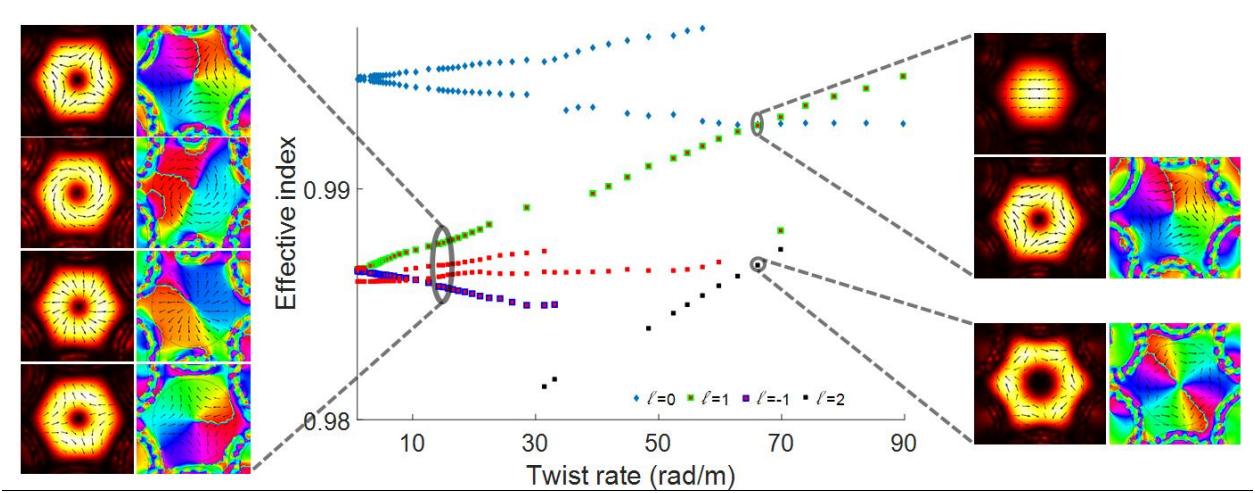

Fig. 8. Simulated supported modes in the fiber as a function of twist rate (2π/twist period) at 0.72 THz. Intensity (modulus of Poynting vector) of modes with respective phase profiles are shown (the black arrows represent the electric field): at 40 cm twist period on the left, at 10 cm twist period on the right. Three groups of modes are highlighted: in blue those that do not lead to OAM, in red those coming from the $LP_{11}$-like modes, with green and blue borders for those leading to a $\ell=\pm 1$ OAM and in black those leading to a $\ell=2$ OAM.

Applying twist removes the degeneracy in the fundamental mode and also in the $LP_{11}$-like modes. The $LP_{11}$ manifold mode splits in 4 non-degenerate modes with a similar distribution to the ones carrying OAM in a standard optical fiber. It is, however, the hybrid ($HE_{21}$ like) modes that carry OAM in this case and the two hybrid modes have opposite rotation direction. At even higher twist rates, some of the modes became very lossy and disappear, and modes with higher order OAM are guided. It should be noted that apparent missing data points arise from coupling of the modes with modes propagating in the outer tubes and therefore having less than 50% of the field square in the core. At a twist period of 10 cm, the one used in the experiments, there are 3 modes guided: the $\ell=0$, and the hybrid modes with $\ell=1$ and $\ell=2$. Moreover, at about this twist rate, the fundamental mode and the OAM mode of order 1 have a refractive index crossing, allowing for efficient coupling from one mode to the other. With an effective index differences in the order of $10^{-3}$ in this frequency band, the coupling length between two modes is in the order of 10 cm, even if very little perturbation is considered. This is consistent with the experiment. Moreover, the structure of the fiber and the twist are not perfect, therefore increasing the coupling between modes. It should be noted that, for these conditions, the ℓ=1 mode is also the mode with the lowest loss (see also Fig. S3(b) in the supplementary material).

Because of this effect, it is reasonable to assume that we are exciting multiple modes or that the excited fundamental mode is transferring energy to the OAM mode and therefore what we observe is a superposition of these. This would explain the asymmetry in the measured OAM mode, where the field of the fundamental mode is in phase with part of the OAM mode and out of phase with the opposite side, leading to constructive interference in one position, destructive interference at the position opposite to it and a gradient in between. To confirm it, we summed the modes obtained from the numerical simulations and we also made a simple analytical model by summing Laguerre-Gaussian modes, the field of which is defined in cylindrical coordinates as a function of space and time as follows (as defined in Ref. 27 but adapted to our time evolution convention):

$$E(r,\phi,z,t) = \sqrt{\frac{2p!}{\pi(p+|\ell|)!}} \frac{1}{w(z)} \left[\frac{r\sqrt{2}}{w(z)}\right]^{|\ell|} e^{\left[\frac{-r^2}{w^2(z)}\right]} L_p^{|\ell|}\left(\frac{2r^2}{w^2(z)}\right) e^{\left[\frac{-kr^2 z}{2(z^2+z_R^2)}\right]} e^{\left[i(2p+|\ell|+1)\tan^{-1}\left(\frac{z}{z_R}\right)\right]} e^{-i(kz-\omega t+\ell\phi)} e^{-\frac{4\ln(2)(t-t_0)^2}{\Delta t^2}}, \quad (2)$$

where $w(z) = w(0)[(z^2 + z_R^2)/z_R^2]^{1/2}$, $w(0)$ is the beam waist, $z_R$ the Rayleigh range, $p$ the radial order (number of radial nodes in the intensity distribution, set to 0 in our calculation), $k = 2\pi/\lambda$ the wave vector, $\lambda$ the wavelength, $\omega = 2\pi c/\lambda$ the wave angular frequency, $c$ the speed of light, $t_0$ the arrival time of the maximum of the pulse, and $\Delta t$ the full width half maximum pulse duration. It should be noted that the minus sign in $e^{-i(kz-\omega t+\ell\phi)}$ is used for consistency with the mathematical formalism used in the data analysis and numerical simulations.

Figure 9 shows the result of adding the ℓ=0 and ℓ=1 modes with a one to one contribution in electric field, as defined in Eq. (2), which gives a very close match to our measurement. Although the result obtained by summing the numerically calculated modes (Fig. 9(a)) is, as expected, more similar to the measurements than the analytical model (Fig. 9(b)), the two results are quite close, indicating that Laguerre-Gaussian modes are a good and mathematically simpler approximation for the system. Confirming the presence of the two modes does not explain whether the OAM carrying mode is generated by directional coupling or already excited at the fiber input. We calculated the mode overlap integral between the excitation and the two modes (for details see the relevant section of the supplementary material) and confirmed that it is unlikely that the ℓ=1 mode could be sufficiently excited at the input in our experiment. This, along with the modal crossing at the measured twist rate, suggests that the OAM mode is indeed the result of mode conversion in the fiber.

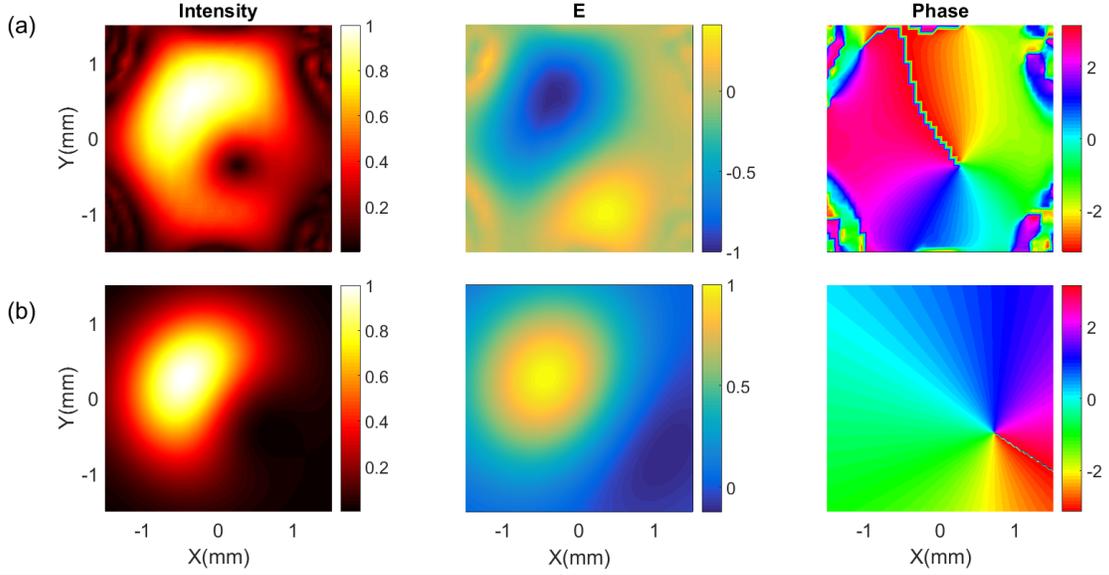

Fig. 9. Intensity profile $|E_x|^2$, electric field, and phase at 0.72 THz calculated by linear combination of the $\ell=0$ and $\ell=1$ modes, both with the same contribution, obtained by: (a) finite elements numerical simulation and (b) analytical Laguerre-Gaussian model, Eq. (2) calculated at $z = 0$ and $t = t_0$.

From the time evolution of the measurement, video of Fig. 6, it is possible to see that the phase singularity of the mode is moving from the bottom right corner towards the middle of the fiber. This is an indication of the combination of the modes not being constant in time. It is possible that either more than two modes contribute to the field, or the two modes are not traveling at the same group velocity. From the COMSOL simulation, and the dependence of the refractive index with frequency of the two modes, it is possible to deduce that the fundamental mode is faster than the OAM mode and therefore arrives earlier. A rough estimation from the simulations leads to a 1-2% difference in velocity. After 10 cm propagation length, this would result in the two modes arriving 1.5 ps time apart. Figure 10 shows the time evolution calculated using the same two modes being Gaussian pulses with a 7 ps duration (FWHM, chosen to be similar to the measured output pulse), separated by 1.5 ps. The match between the simulations and the measurements is remarkable, confirming our interpretation of the fields measured being a superposition of the fundamental and $\ell = 1$ modes.

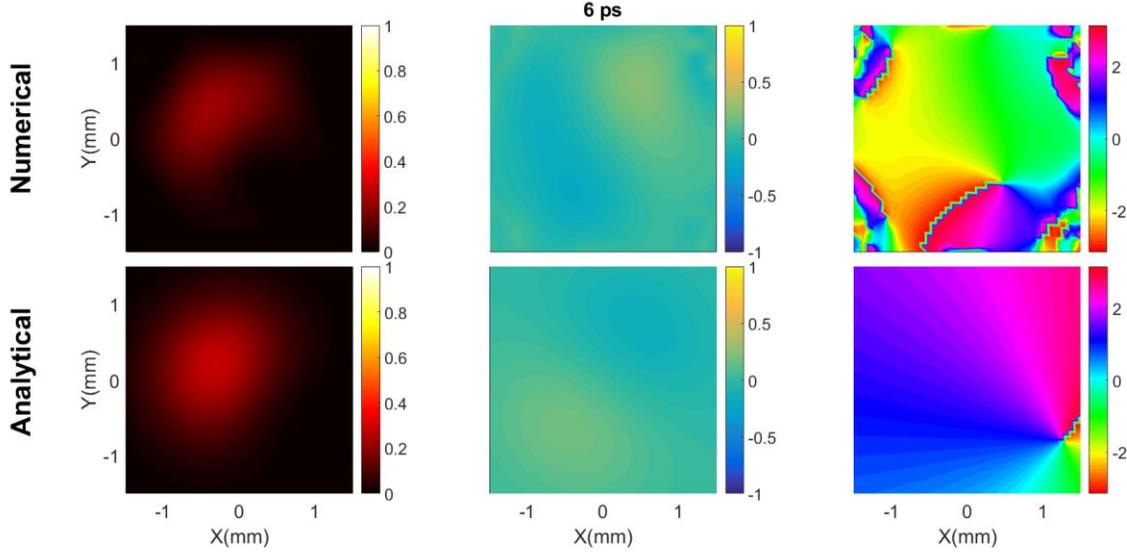

Fig. 10. Time evolution of the intensity profile $|E_x|^2$, electric field, and phase at 0.72 THz generated by analytically adding the numerical simulated (top) and Laguerre-Gaussian (bottom) modes of order 0 and 1 and with the $\ell=0$ mode arriving 1.5 ps before the $\ell=1$ mode (multimedia view).

The time evolution of the electric field component measured (Fig. 6) provides some additional clues as to the modal composition of the fields: positive and negative lobes are spiraling, and visible along the entire spiral. Because of the single polarization measurement, this hints toward the measurement of a field that is not purely radially or azimuthally polarized, but hybrid, which is in agreement with the simulation. Measurement of the complementary linear polarization $E_y$ could help elucidate the exact modal decomposition further.

**CONCLUSION**

In conclusion, we have reported a new flexible hollow waveguide for the THz regime. The waveguide flexibility allowed mechanical twisting to generate modes with OAM in a broad bandwidth (0.6-0.9 THz), starting from a simple Gaussian beam with linear polarization. The extreme flexibility allows to explore twist periods of order several tens of wavelengths, which is difficult in other implementations. The unique features of THz TDS allowed us to measure and therefore visualize the vortex nature of OAM of light. This innovative platform has the potential for changing the panorama of THz waveguides and allowing for a wide range of mode manipulation options within the waveguide. In addition to providing new perspectives for delivery of THz radiation, for generation of OAM or both, the results here reported give access to new physics in the field of twisted fibers by using THz TDS to obtain information on the the interaction of radiation and the structure. Examples are: measuring the split in degeneracy between left and right OAM modes; and measuring phase and group velocity by looking at the velocity of rotation of the electric field and the intensity of the mode, respectively.

Preliminary simulations show that separating the capillaries and therefore the index of the core mode and the capillaries' modes, will result in cleaner and lower loss OAM modes. Some future investigation and applications of this technology include manipulating the polarization: from simple rotation to changing the polarization state; generation of higher order OAM, by higher twist rates or by slightly changing the parameters of the structure. Indeed the trasmission bands where the untwisted fiber supports only the fundamental mode, with a certain twist rate start supporting high order modes with OAM, the order of which is determined by the structure.

## SUPPLEMETARY MATERIAL

See supplementary materials for information about: "fiber flexibility", "transmission of the twisted fiber", "loss of the high order modes" and "mode overlap integral and coupling". The data used for the results here reported are openly available with DOI: 10.5281/zenodo.1164309.

## ACKNOWLEDGMENTS

The authors acknowledge Richard Lwin for providing the polyurethane tubes and Juliano G. Hayashi for fruitful discussions. This work was supported by the Marie Sklodowska-Curie grant of the European Union's Horizon 2020 research and innovation programme (708860) and the Australian Research Council under the Discovery Project scheme number DP170103537.

[45]S. Fleming et al., "Tunable metamaterials fabricated by fiber drawing," J. Opt. Soc. Am. B, vol. 34, no. 7, p. D81, 2017.

[46]H. Li et al., "Guiding terahertz orbital angular momentum beams in multimode Kagome hollow-core fibers," Opt. Lett., vol. 42, no 2, p. 179, 2017.

19